\documentclass[prd,aps,secnumarabic,superscriptaddress,showpacs]{revtex4}
\def\sc{\mbox{\rule[-1pt]{0pt}{11pt}}}
\def\ds{\displaystyle}
\usepackage[dvips]{graphicx,color}
\begin{document}
\title{Comment on ``Damping of Tensor Modes in Cosmology'' }
\author{Duane A. Dicus}
\affiliation{Center for Particle Physics, University of Texas, Austin,TX 78712} \email{dicus@physics.utexas.edu}
\author{Wayne W. Repko}
\affiliation{Department of Physics and Astronomy, Michigan State University, East Lansing, MI 48824} \email{repko@pa.msu.edu}
\date{\today}
\begin{abstract} We provide an analytic solution to the short wave length limit of the integro-differential equation describing the damping of the tensor modes of gravitational waves.
\end{abstract}
\maketitle

In a recent paper \cite{weinberg}, Weinberg gives an integro-differential equation for the propagation of cosmological gravitational waves. In particular he writes an equation for the perturbation to the metric $h_{ij}({\bf\,x},t)$ and then defines $\chi(u)$ as
\begin{equation}\label{chi}
h_{ij}(u)\,=\,h_{ij}(0)\chi(u)\,,
\end{equation}
where $u$ is the conformal time multiplied by the wavenumber
\begin{equation}\label{time}
u\,=\,k\,\int^t\frac{dt'}{a(t')}\,.
\end{equation}
$\chi(u)$ satisfies an integro-differential equation which for short wavelengths (wavelengths which entered the horizon while the universe was still radiation dominated) is given by \cite{weinberg,eq1}
\begin{equation}\label{basiceqn}
u^2\chi''(u)+2u\chi'(u)+u^2\chi(u)=-24f_{\nu}(0)\int_0^{u}dU\,K(u-U)\chi'(U)\,.
\end{equation}
The fraction of the energy density in neutrinos is $f_{\nu}(0)\,=\,0.40523$ and the kernel $K$ will be discussed in detail below. The initial conditions are
\begin{equation}\label{ic}
\chi(0)\,=\,1,\hspace{3mm}\chi'(0)\,=\,0\,.
\end{equation}
In the absence of free-streaming neutrinos the right-hand-side of Eq.(\ref{basiceqn}) is zero and $\chi(u)=\sin(u)/u$.  The suppression of these modes is due to the presence of the neutrinos where the solution of Eq.(\ref{basiceqn}) approaches, for $u>>1$,
\begin{equation}\label{asmpt}
\chi(u)\longrightarrow\,A\,\sin(u+\delta)/u\,,
\end{equation}
and the value of $A^2$ is the quantitative measure of that suppression. After deriving the above Weinberg says ``A numerical solution of Eqs.\,(22) and (23) [our Eqs.(\ref{basiceqn}) and (\ref{ic})] shows that $\chi(u)$ follows the $f_{\nu}=0$ solution pretty accurately until $u\approx1$, when the perturbation enters the horizon, and thereafter rapidly approaches the asymptotic form (Eq.(\ref{asmpt})), with $A=0.8026$ and $\delta$ very small.''

Our purpose here is to provide an analytic solution of Eqs.(\ref{basiceqn}) and (\ref{ic}) so that readers can verify the quoted statement for themselves.

The importance of these results is shown by the remainder of the quote ``This asymptotic form provides the initial condition for the later period when the matter energy density becomes first comparable to and then greater than that of radiation, so the effect of neutrino damping at these later times is still only to reduce the tensor amplitude by the same factor $A=0.8026$.  Hence, for wavelengths that enter the horizon after electron-positron annihilation and well before radiation-matter equality, all quadratic effects of the tensor modes in the cosmic microwave background, such as the tensor contribution to the temperature multipole coefficients $C_{\ell}$ and the whole of the ``B-B'' polarization multipole coefficients $C_{\ell\,B}$, are $35.6\%$ less than they would be without the damping due to free-streaming neutrinos.''

A solution to Eq.\,(\ref{basiceqn}) is a series of spherical Bessel functions \cite{JMS}
\begin{equation}\label{series}
\chi(u)\,=\,\sum_{n=0}^{\infty}\,a_{n}j_{n}(u)\,.
\end{equation}
Inserting Eq.\,(\ref{series}) in the left-hand-side of Eq.(\ref{basiceqn}) and using the differential equation for spherical Bessel functions leaves
\begin{equation}\label{LHS}
\sum_{n=0}^{\infty}n(n+1)a_{n}j_{n}(u)\,.
\end{equation}
The right-hand-side of Eq.\,(\ref{basiceqn}) requires more work.  The kernel is itself a sum of spherical Bessel functions
\begin{eqnarray}
K(u)\,&=&\,\frac{1}{16}\int_{-1}^{1}\,dx\,(1-x^2)^2e^{ixu}   \label{K1}  \\
      &=&\,-\frac{\sin\,u}{u^3}\,-\,\frac{3\,\cos\,u}{u^4}\,+\,\frac{3\,\sin\,u}{u^5}   \label{K2}  \\
      &=&\,\frac{1}{15}\left(j_{0}(u)\,+\,\frac{10}{27}j_{2}(u)\,+\,\frac{3}{7}j_{4}(u)\right),   \label{K3}
\end{eqnarray}
and the derivative of $\chi(u)$ is given by
\begin{eqnarray}
\chi'(u)\,&=&\!\sum_{n=0}^{\infty}a_{n}j'_{n}(u)   \label{der1}  \\
&=&\!\sum_{n=0}^{\infty}a_{n}\frac{\left[nj_{n-1}(u)-(n+1)j_{n+1}(u)\right]}{(2n+1)}\,. \label{der2}
\end{eqnarray}
So the RHS of Eq.\,(\ref{basiceqn}) is $-C\,I(u)$ with $C=1.6\,f_{\nu}(0)=0.648368$ and $I(u)$ given by
\begin{equation}\label{Iu}
I(u)\,=\,\sum_{m=0,2,4}d_{m}\,\sum_{n=0}^{\infty}\frac{a_{n}}{(2n+1)}I_{n,m}(u)
\end{equation}
where
\begin{equation}\label{Inm}
I_{n,m}(u)  = \int_{0}^{u}dU\,j_{m}(u-U)\left[nj_{n-1}(U)-(n+1)j_{n+1}(U)\right]\,.
\end{equation}
The $d_{m}$ are given by Eq.\,(\ref{K3}) where we have factored out a $\frac{1}{15}$. Evaluating $I_{n,m}$ is an exercise in using Abramowitz and Stegun \cite{AS}(AS). First, use the fact that the Fourier transform of a Legendre polynomial is a spherical Bessel function (AS 10.1.14)
\begin{equation}\label{FT}
j_{n}(x)=\frac{(-i)^n}{2}\int_{-1}^{1}\,ds\,e^{ixs}\,P_{n}(s)
\end{equation}
to replace both Bessel functions in Eq.\,(\ref{Inm}).  This makes the integral over $U$ trivial and we have
\begin{eqnarray}\label{st}
I_{n,m}(u) = \frac{(-i)^{n+m}}{4}\int_{-1}^{1}ds\int_{-1}^{1}dt \frac{e^{itu}\,-\,e^{isu}}{t-s}P_{m}(s)[nP_{n-1}(t)\,+\,(n+1)P_{n+1}(t)]\,.
\end{eqnarray}
Now use the definition of the Legendre function of the second kind,
\begin{equation}\label{AS883}
Q_{n}(z)=\frac{1}{2}\int_{-1}^{1}\,dx\,(z-x)^{-1}P_{n}(x),
\end{equation}
(AS 8.8.3), to evaluate the integral in Eq.\,(\ref{st}) over the variable that does not appear in the exponent (s in the first term, t in the second) to obtain
\begin{equation}\label{Inm1}
I_{n,m}(u)=\frac{(-i)^{n+m}}{2}\int_{-1}^{1}dte^{itu}\left\{Q_{m}(t) [nP_{n-1}(t)+(n+1)P_{n+1}(t)]+P_{m}(t)[nQ_{n-1}(t)+ (n+1)Q_{n+1}(t)] \right\}\,.
\end{equation}
Next, by replacing the remaining exponential with the familiar expression from quantum mechanical scattering,
\begin{equation}\label{QM}
e^{itu}\,=\,\sum_{\ell}(2\ell+1)i^{\ell}j_{\ell}(u)P_{\ell}(t)\,,
\end{equation}
the expression for $I_{n,m}(u)$ becomes
\begin{eqnarray}\label{PQ1}
\lefteqn{I_{n,m}(u)\,=\,\sum_{\ell}\frac{(2\ell+1)}{2}(-i)^{n+m-\ell}j_{\ell}(u)\left\{\int_{-1}^{1}dt P_{\ell}(t)Q_{m}(t)[nP_{n-1}(t)+(n+1)P_{n+1}(t)]\right.}\hspace{1.5in}  \nonumber  \\
&   &\,+\,\left.\int_{-1}^{1}dtP_{\ell}(t)P_{m}(t) [nQ_{n-1}(t)+(n+1)Q_{n+1}(t)]\right\}\,.
\end{eqnarray}
Eq.\,(\ref{PQ1}) can be simplified using (AS 8.6.19),
\begin{equation}\label{AS8619}
Q_{m}(x)=\frac{1}{2}P_{m}(x)\ln\frac{1+x}{1-x}\,-\,W_{m-1}(x)\,,
\end{equation}
where
\begin{equation}\label{W}
W_{m-1}(x)=\!\!\sum_{k=0}^{[(m-1)/2\,]}\frac{2m-4k-1}{(2k+1)(m-k)}P_{m-2k-1}(x)\,,
\end{equation} and the formula
\begin{equation}\label{PP}
P_{\ell}(x)P_{m}(x)=\!\!\!\!\sum_{L=|\ell-m|}^{\ell+m}\!\!\!|\!<\ell,0,m,0|L,0>\!|^2P_{L}(x)
\end{equation}
to express $P_\ell(x)P_m(x)$ in terms of $P_L(x)$'s and, with the aid of Eq.\,(\ref{AS8619}), $P_\ell(x)Q_m(x)$ in terms of $Q_L(x)$'s as
\begin{equation}\label{PQ}
P_{\ell}(x)Q_{m}(x)=\sum_{L=|\ell-m|}^{\ell+m}\Big[|<\ell,0,m,0|L,0>|^2 \left(Q_{L}(x)+W_{L-1}(x)\right)\Big]- P_{\ell}(x)W_{m-1}(x)\,.
\end{equation}
Finally, the terms in Eq.\,(\ref{PQ1}) involving the products of $P_n(x)$'s and $Q_m(x)$'s cancel using (AS 8.14.10),
\begin{equation}\label{PQm}
\int_{-1}^{1}dx\left(Q_{L}(x)P_{m\pm 1}(x)+P_{L}(x)Q_{m\pm 1}(x)\right)=0\,,
\end{equation}
and $I_{n,m}(u)$ reduces to
\begin{eqnarray}\label{final}
I_{n,m}(u)&=&\sum_{\ell}\frac{2\ell+1}{2}(-i)^{n+m-\ell}j_{\ell}(u)
\Big\{\int_{-1}^{1}\!dt\!\!\!\sum_{L=|\ell-m|}^{\ell+m}\!\!|<\ell,0,m,0|L,0>|^2W_{L-1}(t) \left[nP_{n-1}(t)+(n+1)P_{n+1}(t)\right]  \nonumber  \\
&  & \mbox{}-\,\int_{-1}^{1}dtP_{\ell}(t)W_{m-1}(t) [nP_{n-1}(t)+(n+1)P_{n+1}(t)]\Big\}
\end{eqnarray}
The contributions to the coefficient of each $j_{\ell}(u)$ in Eq.\,(\ref{final}) can be straight-forwardly evaluated; those in the sum by directly using orthogonality and the remaining terms by expressing the product of two $P_\ell$'s as a sum of $P_\ell$'s and again using orthogonality. The orthogonality of the Legendre functions means that the $\ell$ which is summed over in Eq.\,(\ref{final}) can only take on the values $n+2k$ where $k=0,1,2,\cdots$ so we replace
$$
\sum_{m=0,2,4}\frac{d_{m}}{2n+1}I_{n,m}(u)   \nonumber
$$
with
$$
\sum_{k=0}^{\infty}c_{n,k}j_{n+2k}(u)\,,   \nonumber
$$
i.e., the sum over $\ell$ in $I_{n,m}(u)$ is replaced by a sum over $k$ and each $c_{n,k}$ is the sum of the contributions from the three terms in the kernel, $m=0,2,4$. Setting Eq.\,(\ref{LHS}) equal to $-\,CI(u)$ we have
\begin{equation}\label{cnm}
\sum_{n=0}^{\infty}n(n+1)a_{n}j_{n}(u)=-C\sum_{n,k=0}^{\infty} a_{n}c_{n,k}j_{n+2k}(u)\,,
\end{equation}
where the $c_{n,k}$ are known numbers and we can find the expansion coefficients, $a_{n}$, recursively by equating the coefficients of each order Bessel function in Eq.\,(\ref{cnm}).

The coefficients of $j_1(u)$ in Eq.\,(\ref{cnm}) give
\begin{equation}\label{a1}
2\,a_1\,=\,-Cc_{1,0}a_1\,,
\end{equation}
where $c_{n,0}$ is equal to $1\,-\,\delta_{n,0}$ so $c_{1,0}$ is $1$.  The only solution of this equation is $a_1=0$. This ensures that the second of the initial conditions, Eq.\,(\ref{ic}), is satisfied. The equality of the coefficient of $j_3(u)$ shows that $a_3$ is proportional to $a_1$.  Similarly $a_5$ is a linear combination of $a_1$ and $a_3$, $a_7$ a linear combination of $a_1$, $a_3$, and $a_5,\,\cdots$. Thus the coefficients of all the odd order Bessel functions in Eq.(\ref{series}) are zero. There is no mixing between the coefficients of the odd order Bessel functions and those of even order because the Clebsch-Gordan coefficients $<a,0,b,0|c,0>$ are zero if $a+b+c$ is an odd number.
\begin{table}[h]
\begin{center}
\begin{tabular}{| c | ccccccc |}
\hline
$n$ & $c_{n,0}$ & $c_{n,2}$ & $c_{n,4}$ & $c_{n,6}$ & $c_{n,8}$ & $c_{n,10}$ & $c_{n,12}$  \\ [2mm] \hline
\mbox{\rule{0pt}{14pt}}$0$ & $0$ & $\ds-\frac{5}{2}$ & $\ds-\frac{3}{2}$ & $\ds-\frac{13}{60}$ & $\ds\frac{17}{735}$ & $\ds-\frac{1}{168}$ & $\ds\frac{125}{58212}$  \\ [3mm]
$2$ & $1$ & $\ds-\frac{3}{4}$ & $\ds-\frac{13}{21}$ & $\ds-\frac{17}{168}$ & $\ds\frac{1}{84}$ & $\ds-\frac{5}{1512}$ &  \\ [3mm]
$4$ & $1$ & $\ds-\frac{143}{210}$ & $\ds-\frac{221}{420}$ & $\ds-\frac{41}{504}$ & $\ds\frac{425}{45738}$ &  &  \\ [3mm]
$6$ & $1$ & $\ds-\frac{255}{392}$ & $\ds-\frac{17}{35}$ & $\ds-\frac{93575}{1280664}$ & & &  \\ [3mm]
$8$ & $1$ & $\ds-\frac{19}{30}$ & $\ds-\frac{25}{54}$ & & & & \\ [3mm]
$10$ & $1$ & $\ds-\frac{575}{924}$ & & & & & \\ [3mm]
$12$ & $1$ & & & & & & \\
\hline
\end{tabular}
\caption{The coefficients needed for Eq.\,(\ref{cnm}) to evaluate Eq.\,(\ref{series}) up to $n=12$. \label{coef}}
\end{center}
\end{table}

Thus the only nonzero $a_{n}$ in Eq.\,(\ref{series}) are those with even $n$. $a_{0}$ does not appear in Eq.\,(\ref{cnm}) but is determined by the first of the initial conditions of Eq.\,(\ref{ic}), which fixes it to be unity. The $c_{n,k}$ necessary to find $a_{2},\cdots,a_{12}$ are shown in Table \ref{coef}.

The equations for these $a_{n}$ can be read off from Eq.\,(\ref{cnm}). Using $a_{0}=1$, we have
\begin{equation}
\begin{array}{ccc}
a_{2}\,&=&\,-C\,\frac{\sc\ds c_{0,2}}{\sc\ds 6+C}    \\
a_{4}\,&=&\,-C\,\frac{\sc\ds c_{0,4}+a_{2}c_{2,2}}{\sc\ds 20+C} \\
\vdots & & \\
a_{2n}\,&=&-\,\frac{\sc\ds C\sum_{k=0}^{n-1}a_{2k}c_{2k,2n-2k}}{\sc\ds 2n(2n+1)+C}
\end{array}\,.
\end{equation}

\begin{figure}[h]
\centering
\includegraphics[width=2.75in]{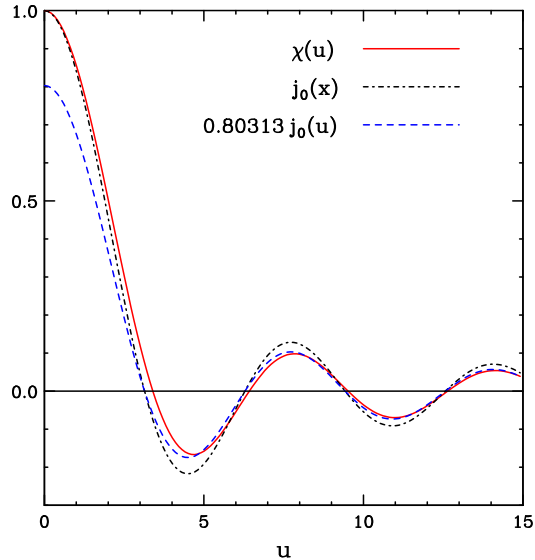}
\caption{(color online) The solution $\chi(u)$ to Eq.\,(\ref{basiceqn}) (solid) is compared to the $f_\nu=0$ solution $j_0(u)$ (dotdashed) and the asymptotic solution $0.80313\,j_0(u)$ (dashed). \label{chiall}}
\end{figure}

Numerically we find that the $a_{n}$ decrease very quickly
\begin{eqnarray*}
a_0\,&=&\,1.0  \\
a_2\,&=&\,0.243807  \\
a_4\,&=&\,5.28424\times10^{-2}  \\
a_6\,&=&\,6.13545\times10^{-3}  \\
a_8\,&=&\,2.97534\times10^{-4}  \\
a_{10}\,&=&\,6.16273\times10^{-5}  \\
a_{12}\,&=&\,-4.78885\times10^{-6}\,.
\end{eqnarray*}

\vspace*{24pt} For large argument, all of the even order Bessel functions go as $\pm\sin\,x/x$ so the $A$ in Eq.\,(\ref{asmpt}) is
\begin{eqnarray}
A\,&=&\,\sum_{n=0}^{5}(-1)^{n}a_{2n}\,,  \nonumber  \\
   &=&\,0.80313\,.
\end{eqnarray}
Since there are no odd order Bessel functions in the expansion, the phase $\delta$ in Eq.\,(\ref{asmpt}) is zero.

The extent to which $\chi(u)$ departs from the the $f_\nu(0)=0$ solution $j_0(u)$ and approaches the asymptotic solution $0.80313\,j_0(u)$ is illustrated in Fig.\ref{chiall} We see that $\chi(u)$ is discernably different from $j_0(u)$ by $u=1.5-2$ and that it is essentially identical to $0.80313\,j_0(u)$ by $u=5$.

\begin{acknowledgements}
We wish to thank S. Weinberg for his comments and S. Radford for a careful reading of the manuscript. This research was supported in part by the National Science Foundation under Grant PHY-0274789 and by the United States Department of Energy under Contract No. DE-FG03-93ER40757.
\end{acknowledgements}

\end{document}